\documentclass[5p]{elsarticle}
\usepackage{graphicx}

\title{Spontaneous aggregation and global polar ordering in squirmer suspensions\tnoteref{r1}}
\author{F. Alarc\'on \corref{cor2}}
\author{I. Pagonabarraga \corref{cor1}}
\cortext[cor2]{falarcon@ffn.ub.es}
\cortext[cor1]{ipagonabarraga@ub.edu}
\address{Departament de F\'isica Fonamental, Universitat de Barcelona,\\ C. Mart\'i i Franqu\'es, 1, 08028 Barcelona, Spain.
}
\begin{document}
\begin{abstract}
We have developed numerical simulations of three dimension suspensions of active particles  to characterize the capabilities of  the hydrodynamic stresses induced by active swimmers to promote global order and  emergent structures in active suspensions. We have considered squirmer suspensions  embedded in a fluid modeled under a Lattice Boltzmann scheme. We have found that  active stresses play a central role  to decorrelate the collective motion of squirmers and that  contractile squirmers develop significant aggregates.
\end{abstract}
\maketitle

\section{Introduction}
 Collective motion can be observed at a variety of scales, ranging from herds of large   to  bacteria colonies or   the active motion of organelles inside cells. Despite the long standing interest of the wide implications of  collective motion in biology, engineering and  medicine (as for example, the ethological implications of the signals exchanged between  moving animals, the evolutionary benefits of moving in groups for individuals and for species, the design of robots which can accomplish a cooperative tasks without central control, the understanding of   tumor growth or wound healing to mention a few), only recently there has been a growing interest in characterizing such global behavior from a statistical mechanics perspective~\cite{ramaswamy_review}.

Although  a variety ingredients and mechanisms have been reported to describe the  signaling and cooperation among individuals which move collectively, it is important to understand the underlying, basic physical principles that can provide simple means of cooperation and can  lead to emerging patterns and structures~\cite{dion}. We want to analyze the capabilities of basic physical  ingredients to generate  emerging structures in  active particles which self propel in an embedding fluid medium. 
These systems constitute an example of active fluids, systems which generate stresses by the conversion of chemical into mechanical energy. To this end, we will consider model suspensions of swimming particles (building on the squirmer model introduced by Lighthill~\cite{Lighthill}) and will analyze   a hydrodynamically-controled route to flocking. 
We will use a hybrid description of an active suspension, which combines the individual dynamics of   spherical swimmers with a kinetic model for the solvent. We can identify the emergence of global orientational order and correlate it with the formation of spontaneous structures where squirmers aggregate and form flocks of  entities that swim along together. This simplified approach allows us to identify the role of active stresses and self-propulsion to lead both to global orientational order and  aggregate formation. Even if in real systems other factors can also control the interaction and collective behaviors of active suspensions, the present description shows that hydrodynamics itself is enough to  promote cooperation in these systems which are intrinsically out of equilibrium.

This work is organized as follows. In section 2.1 we present the theoretical frame of the simulation technique that we have applied, while in section 2.2 we describe the squirmer model that we have used and introduce the relevant parameters which characterize its hydrodynamic behavior  and in section 2.3 we give a detailed explanation of the simulation parameters and the systems we have studied. Section 3 is devoted to analyze the global polar order parameter   and to study quantitatively the orientation that squirmer suspensions display. In Section 4  flocking is studied via  generalized radial distribution functions, moreover to characterize the time evolution of the formed flocks, we calculated the time correlation function of the density fluctuations, the results are shown in this section also.
% temporal correlation of density fluctuations.
We conclude in Section 5 indicating the main results and their  implications.

%%%%%%%%%%%%%%%%%%%%%%%%%%%%%%%%%%%%%%   
\section{Theoretical Model}
%%%%%%%%%
\subsection{Lattice Boltzmann Scheme}
We consider a model for microswimmer suspensions composed by spherical  particles embedded in a fluid. The fluid   is modeled using a Lattice Boltzmann approach. Accordingly,  the solvent is described in terms of a distribution function $f_i(\vec{r};t)$ in each node of the lattice. The distribution function evolves at discrete time steps, $\Delta t$,  following the lattice Boltzmann equation (LBE):
\begin{eqnarray}\label{LBE}
f_i\left(\vec{r}+\vec{c_i}\Delta t, t+\Delta t\right)=
f_i\left(\vec{r}; t\right)+ \nonumber \\
\Omega _{ij}\left(f_{j}^{eq}\left(\vec{r}; t\right) - 
 f_j\left(\vec{r}; t\right)\right).
\end{eqnarray}
that can be regarded as  the space and time discretized analog of the Boltzmann equation. It includes both the streaming to the neighbouring nodes, which corresponds to the advection of the fluid due to its own velocity, and the relaxation toward a prescribed equilibrium distribution function $f_{j}^{eq}$. This relaxation is determined by the linear collision operator  $\Omega _{ij}$\cite{Succi, scaling, Llopis_epl06}. It corresponds to linearizing the  collision operator of the  Boltzmann equation. If $\Omega _{ij}$ has one single eigenvalue, the method corresponds to the  kinetic  model introduced by  Bhatnagar-Gross-Crook (BGK)~\cite{BGK_92}. The LBE satisfies the Navier-Stokes equations at large scales. In all our simulations we use units such that the mass of the nodes, the lattice spacing and the time step $\Delta t$ are unity and the viscosity is $1/2$, the lattice geometry that we have used was a cubic lattice with 19 allowed velocities, better known as $D_3Q_{19}$ scheme \cite{scaling}. 

The linearity and locality of LBE makes it a useful method to study the dynamic of fluids under complex geometries, as is the case when dealing with particulate suspensions.  Using the distribution function as the central dynamic quantity makes it possible to express the fluid/solid boundary conditions as local rules. Hence, stick boundary conditions can be enforced through  bounce-back of the distribution, $f_i(\vec{r};t)$, on the links joining fluid nodes and lattice nodes inside the  shell which defines the solid particles, also known as boundary links~\cite{Ladd_BBL}.  A microswimmer is modeled as  a spherical shell larger than the lattice spacing. Following the standard procedure, the microswimmer is represented by the  boundary links   which define its surface. Accounting for  the cumulative bounce back of all boundary links allows to extract the net force and torque acting on the suspended particle~\cite{Nguyen_BBL}. The particle dynamics can then be described individually   and particles do not overlap due to a repulsive, short-range interaction among them, given by   
\begin{equation}
v^{ss}\left(r\right)=\epsilon\left(\sigma/r\right)^{\nu_0},
\end{equation}
where $\epsilon$ is the energy scale, and $\sigma$ the characteristic width. The steepness of the potential is set by the exponent $\nu_0$. In all cases we have used $\epsilon = 1.0$, $\sigma = 0.5$ and $\nu_0=2.0$.
%%%%%%%%%%
\subsection{Squirmer Model.} \label{SquirmerMod}

We follow the model   proposed by Lighthill~\cite{Lighthill}, subsequently improved by Blake~\cite{Blake}, for ciliated microorganisms. In this approach, appropriate boundary conditions to the Stokes equation on the surface of the spherical  particles (of radius $R$) are imposed to induce a  slip velocity between the fluid and the particles. This slip velocity determines how the particle can displace in the embedding solvent in the absence of a net force or torque. For axisymmetric motion of a spherical swimmer, the radial, $v_r$ and tangential, $v_{\theta}$ components of the slip velocity can be generically expressed as 
\begin{eqnarray}\label{Gral_swimm}
v_r|_{r_1=R} = \sum_{n=0}^\infty A_n\left(t\right)P_n\left(\frac{\textbf{e}_1 \cdot \textbf{r}_1}{R}\right), \nonumber \\
v_\theta|_{r_1=R} = \sum_{n=0}^\infty B_n\left(t\right)V_n\left(\frac{\textbf{e}_1 \cdot \textbf{r}_1}{R}\right),
\end{eqnarray}n-th
at the squirmer spherical surface, where $P_n$ stands for the $n$-th order Legendre polynomial and $V_n$ is define as
\begin{equation}
V_n\left(\cos \theta \right)= \frac{2}{n(n+1)}\sin \theta~ P'_n(\cos \theta),
\end{equation}
$\textbf{e} _1$ describes the intrinsic director, which moves rigidly with the particle and determines the direction along which a single squirmer will displace, while  $\textbf{r} _1$ represents the position vector  with respect to the squirmer's center, which is always pointing the particle surface and thus $|\textbf{r}_1 | = R$. 
Since the squirmer is moving in an inertialess media, the velocity ${\bf u}$ and pressure $p$ of the fluid are given by the Stokes and continuity equations
\begin{equation}\label{Stokes}
\nabla p = \nu \nabla^2 \bf{u}, \nonumber\\
\nabla \cdot \bf{u} = 0.
\end{equation}
The velocity field generated by a squirmer is the solution of these equations (\ref{Stokes}) under the boundary conditions specified by the slip velocity in the surface of its body, eq.~(\ref{Gral_swimm}).
We will disregard the radial changes of the squirming motion, and will consider $A_n=0$, to focus on a simple model that captures the relevant hydrodynamic features associated to squirmer swimming. Accordingly, we will also disregard the time dependence of the coefficients $B_n$ and will focus on the mean velocity of a squirmer during a beating period~\cite{Llopis_10}. Hence,  from the solution of eqs. ~(\ref{Stokes}) using the slip velocity as a boundary condition (eq.~(\ref{Gral_swimm})), we can write the mean fluid flow induced by a minimal squirmer as
 \begin{eqnarray}
 \textbf{u}_1\left( \textbf{r}_1\right)= -\frac{1}{3} \frac{R^3}{r_1^3} B_1\textbf{e}_1 + B_1\frac{R^3}{r_1^3} \textbf{e}_1 \cdot \textbf{\^r}_1 \textbf{\^r}_1  - \nonumber \\
  \frac{R^2}{r_1^2}B_2P_2\left(\textbf{e}_1 \cdot \textbf{\^r}_1 \right)\textbf{\^r}_1,
 \end{eqnarray}
where we have taken $B_n = 0$, $n > 2$, keeping only the first two terms in the general expression for the slip velocity, Eq.(~\ref{Gral_swimm}). The two non-vanishing  terms account for the leading dynamics effects associates to the squirmers.  While $B_1$ determines the squirmer velocity, along ${\bf e}_1$, and controls its polarity, $B_2$  stands for  the apolar stresses that are generated by the surface waves~\cite{Ishikawa}. The dimensionless parameter $\beta \equiv B_2/B_1$ quantifies the relative relevance of   apolar stresses against  squirmer polarity.   The sign of $\beta$ (determined by that of $B_2$) classifies  contractile squirmers  ( or pullers) with $\beta > 0$ and  extensile squirmers (or pushers) when  $\beta < 0 $. The limiting case when $B_1=0$ corresponds to completely  apolar squirmers (or shakers~\cite{Ramachandran}) which induce fluid motion around them without  self-propulsion. The opposite situation, when $B_2 =0$ corresponds to completely polar,  self-propelling,  squirmers  which do not generate active stresses around them. We will disregard thermal fluctuations;  therefore $B_1$ and $B_2$ are the two parameters which completely characterize  squirmer motion.

\subsection{Simulation Details.}

\begin{figure}[!ht]
 \begin{center}
%    \begin{tabular}{cc} 
%\resizebox{4.5cm\textit{a} }{!}{\includegraphics{gr_slide23Layer.pdf}}
%\scalebox{0.14}{\includegraphics{Swimmers_Init_ALLGND.eps}} 
\scalebox{0.2217}{\includegraphics{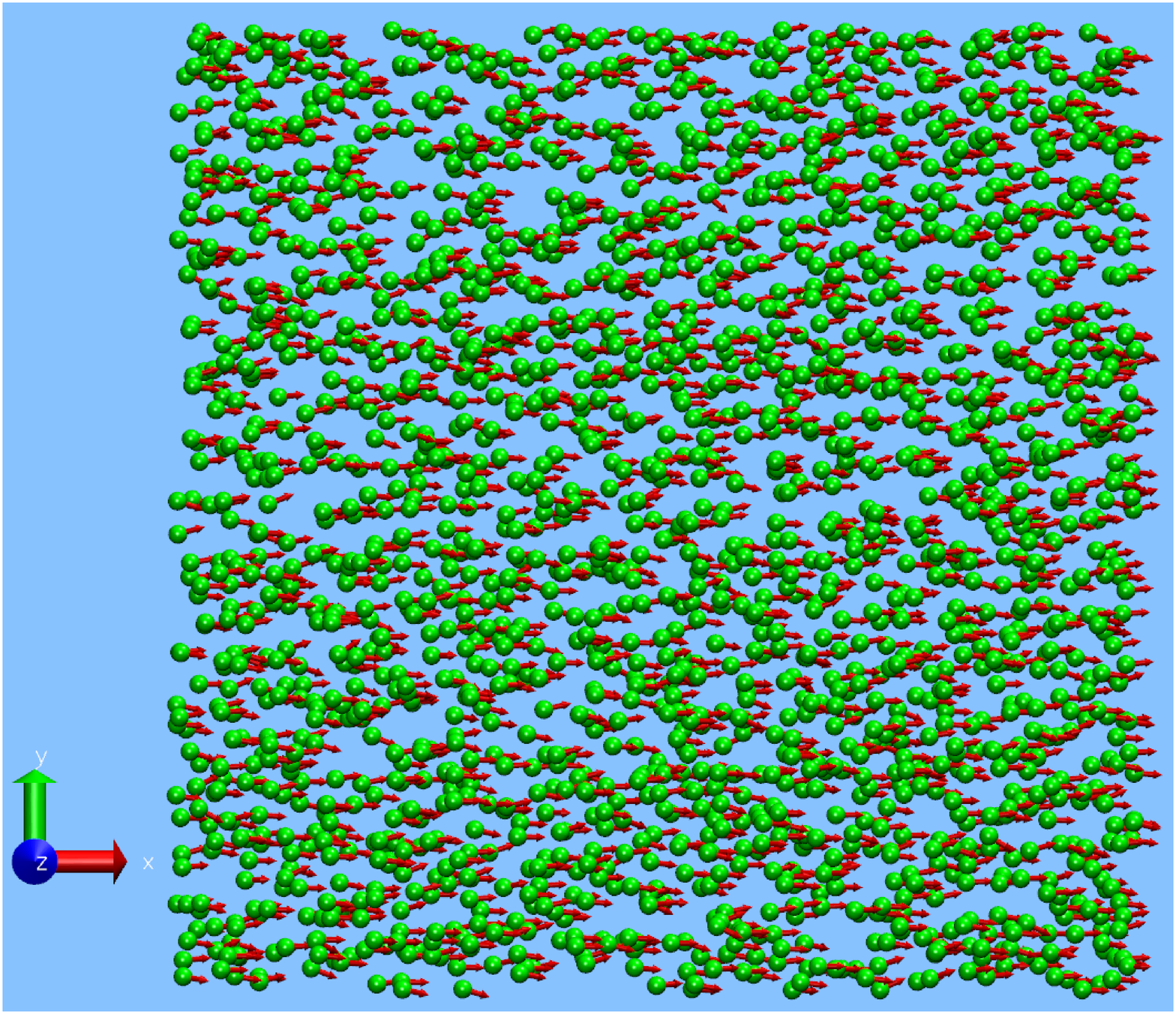}}
 \scalebox{0.2154}{\includegraphics{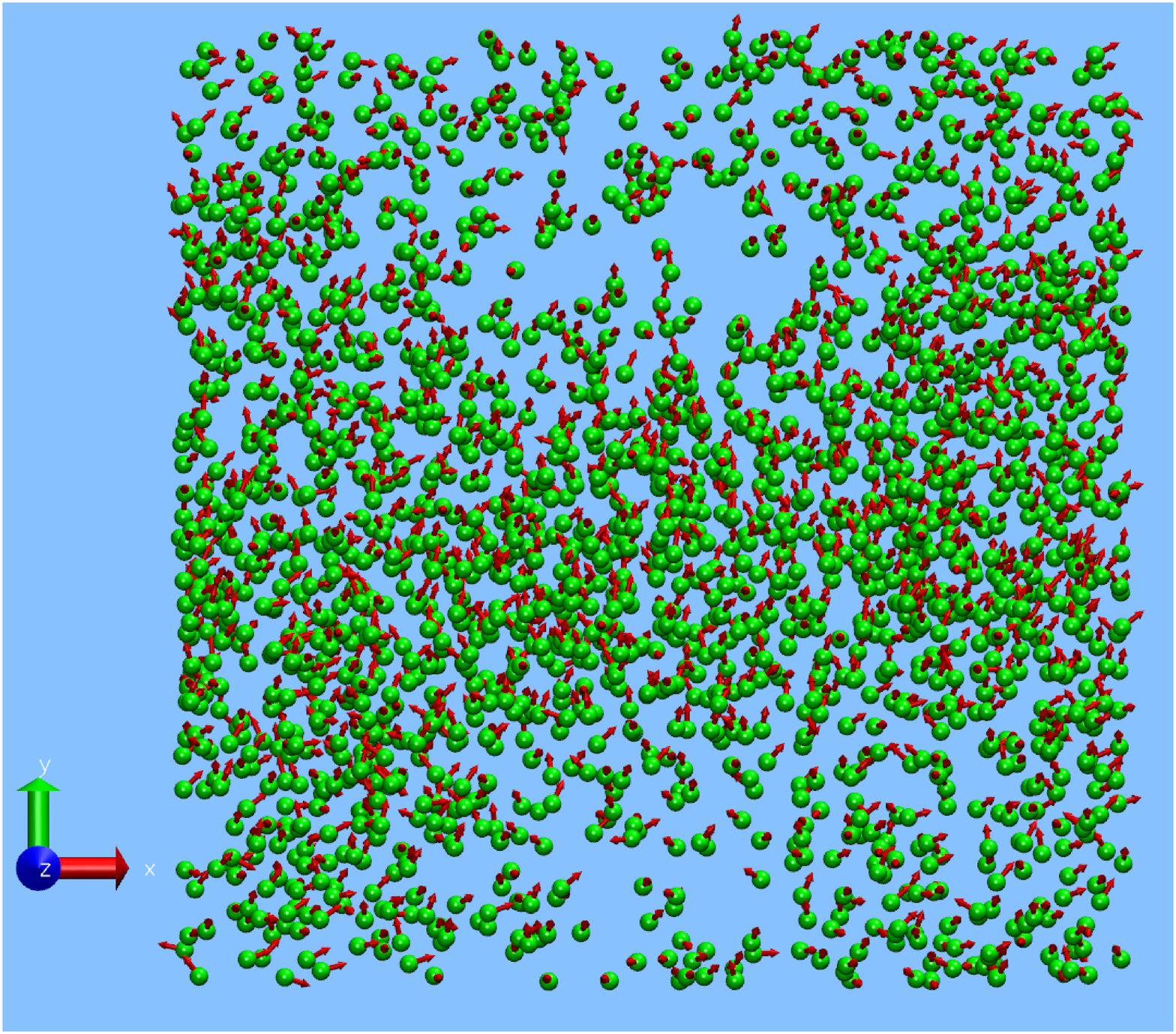}}
 \scalebox{0.22332}{\includegraphics{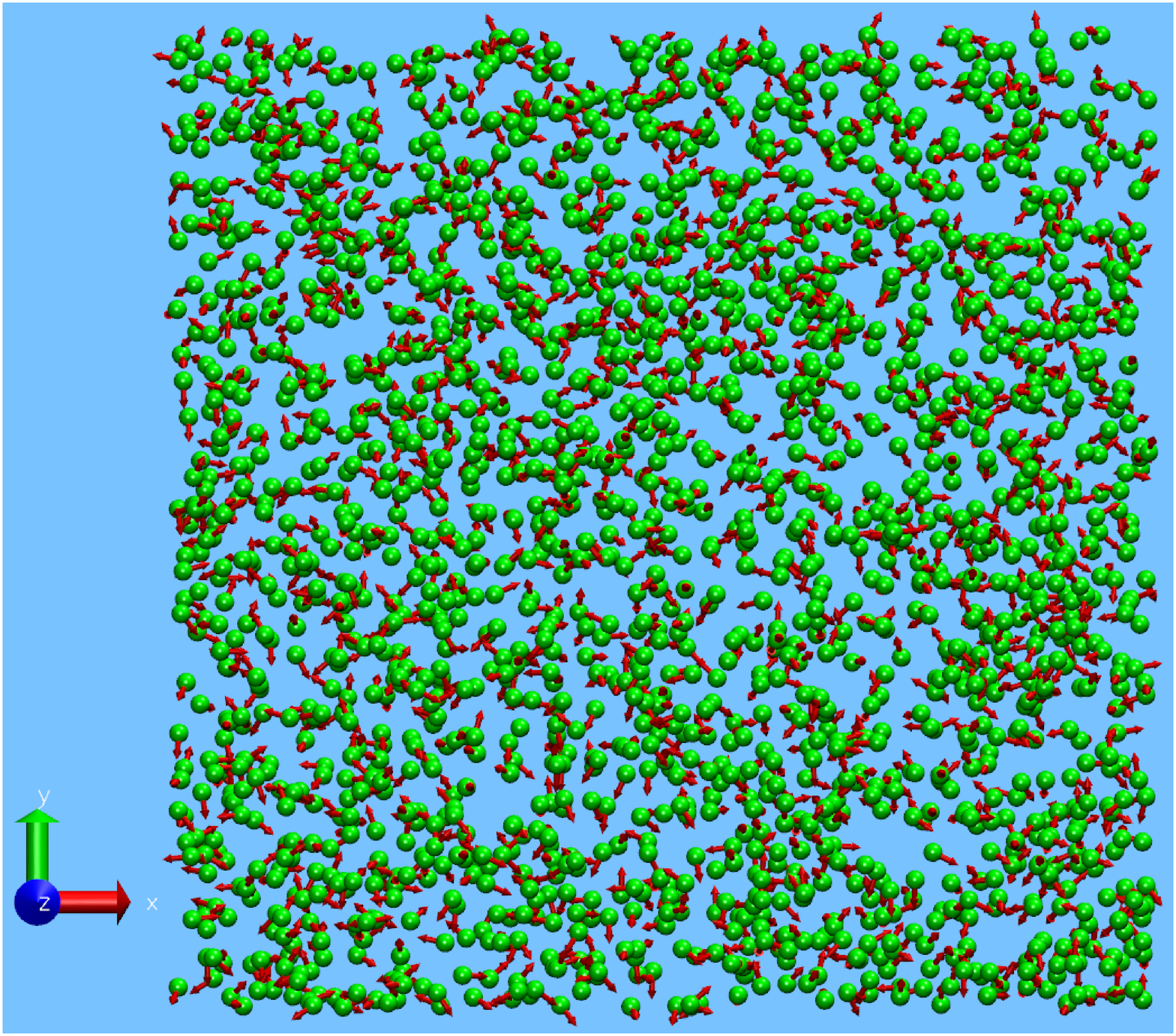}}
%     \includegraphics[scale=0.5]{gr_slide23EPS.pdf}
%    \end{tabular}
\end{center}\vspace{-0.7cm}
  \caption{\small{Snapshots of a simulation with $\beta = 0$ up, $\beta = 0.5$ middle and $\beta = -0.5$ down, at $t/t_0 = 870$. The snapshots have been done using the VMD software \cite{VMD} with the Normal Mode Wizard (NMWiz) plugin \cite{NMWiz}.}}\label{Fotos_3Swimmers}
\end{figure}

All the results that we will discuss  correspond to numerical simulations consisting of $N$ identical spherical particles in a cubic box of volume $L^3$ with periodic boundary conditions. In all cases we have considered $N = 2000$, $R=2.3$ and $L = 100$ (expressed in terms of the lattice spacing). This corresponds to a volume  fraction  $ \phi = 4 \pi N R^3 / (3L^3)=1/10$, 
with a kinematic viscosity of $\nu = 1/2$ (in lattice units)~\cite{stratford}. 
As we will analyze subsequently, active stresses play a significant role in the structures  that squirmers develop when swimming collectively. In Fig.~ \ref{Fotos_3Swimmers}  we compare  characteristic configurations of suspensions for completely polar, contractile and extensile squirmers.  Apolar stresses favor  fluctuations in the squirmer concentration and for  contractile squirmers  there is a clear tendency to form transient, but marked, aggregates. The figure also shows that one needs to distinguish between how squirmers align to swim together and how do they distribute spatially. In the following section we will analyze how active stresses interact with self-propulsion to  affect both aspects of  collective swimming.
%FIGURE%%%%%%
%%%%%%%%%%%%%%%%%%%%%

%%%%%%%%%%%%%%%%%%%%%%%%%%%%%%%%%%%%%%%%%%%%%%%%%%%%%%
\section{Polar Order Parameter.}

\begin{figure}[!ht]
 \begin{center}
%    \begin{tabular}{cc} 
%\resizebox{4.5cm\textit{a} }{!}{\includegraphics{gr_slide23Layer.pdf}}
\scalebox{0.45}{\includegraphics{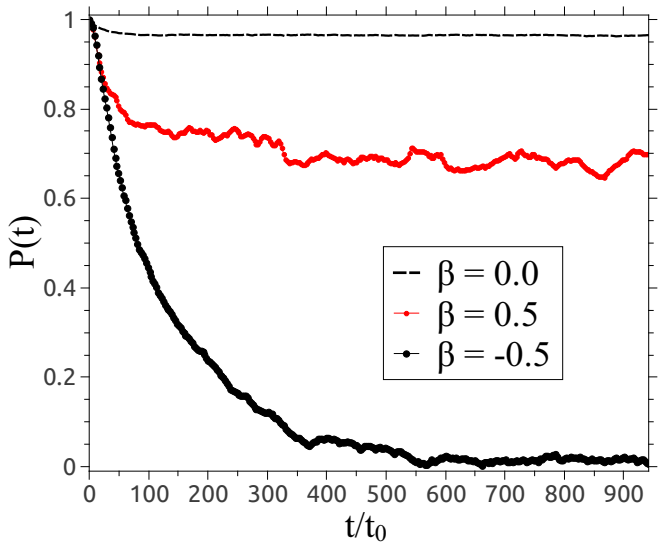}}
%     \includegraphics[scale=0.5]{gr_slide23EPS.pdf}
%    \end{tabular}
\end{center}\vspace{-0.7cm}
  \caption{\small{Polar order parameter $P(t)$, for completely polar squirmers ($\beta = 0$), pullers ($\beta = 0.5$) and pushers ($\beta = -0.5$) initially aligned $P(0) = 1$ and homogeneously distributed in space.}}\label{Pt_algnd}
\end{figure}

In order to quantify the degree of ordering associated to collective squirmer motion, we have computed the global  polar order parameter (eq. \ref{Pt})~ \citep{Lauga}, expressed in terms  of the squirmer intrinsic orientation  $\textbf{e}$, which determines the  direction of swimming for isolated squirmers,
\begin{eqnarray}\label{Pt}
P\left(t\right)=\frac{\mid \sum_{i}^{N}\textbf{e}_i \mid}{N}
\end{eqnarray}
In Fig.~\ref{Pt_algnd} we show the temporal evolution of $P(t)$ as a function of time for completely polar, contractile and extensile  suspensions. The time is normalized by $t_0$ which is the time that a single squirmer needs to self-propel a distance of one diameter, $t_0 \equiv 2R/(2/3~B_1) = 3R/B_1$   The three suspensions start from  a completely aligned initial configuration where squirmers are homogeneously distributed spatially. This figure shows clearly that  squirmers  relax from the given initial configuration to the appropriate steady state and that active stresses have a profound impact on the  ability of squirmers to swim together.  The limiting situation of completely polar swimmers, $\beta=0$, keeps almost perfect ordering. This is because the irrotational flow generated by the translational velocity of the particles is strong enough to maintain a symmetrical distortion in the fluid. Hence, a value of $P(t)$ close to one indicates high  polarity. The other two curves, corresponding to extensile ($\beta =-1/2$) and contractile squirmers ($\beta =1/2$) , indicate that  active stresses generically decorrelate  squirmer motion due to the coupling of the intrinsic direction of squirmer self-propulsion with the local vorticity field induced by the active stresses generated by neighbouring squirmers. However, we do observe a clear difference because  extensile squirmers have completely lost their common degree of swimming while contractile ones still conserve a partial degree of global coherence. 

\begin{figure}[!ht]
 \begin{center}
%    \begin{tabular}{cc} 
%\resizebox{4.5cm\textit{a} }{!}{\includegraphics{gr_slide23Layer.pdf}}
\scalebox{0.30}{\includegraphics[width=30cm,height=25cm]{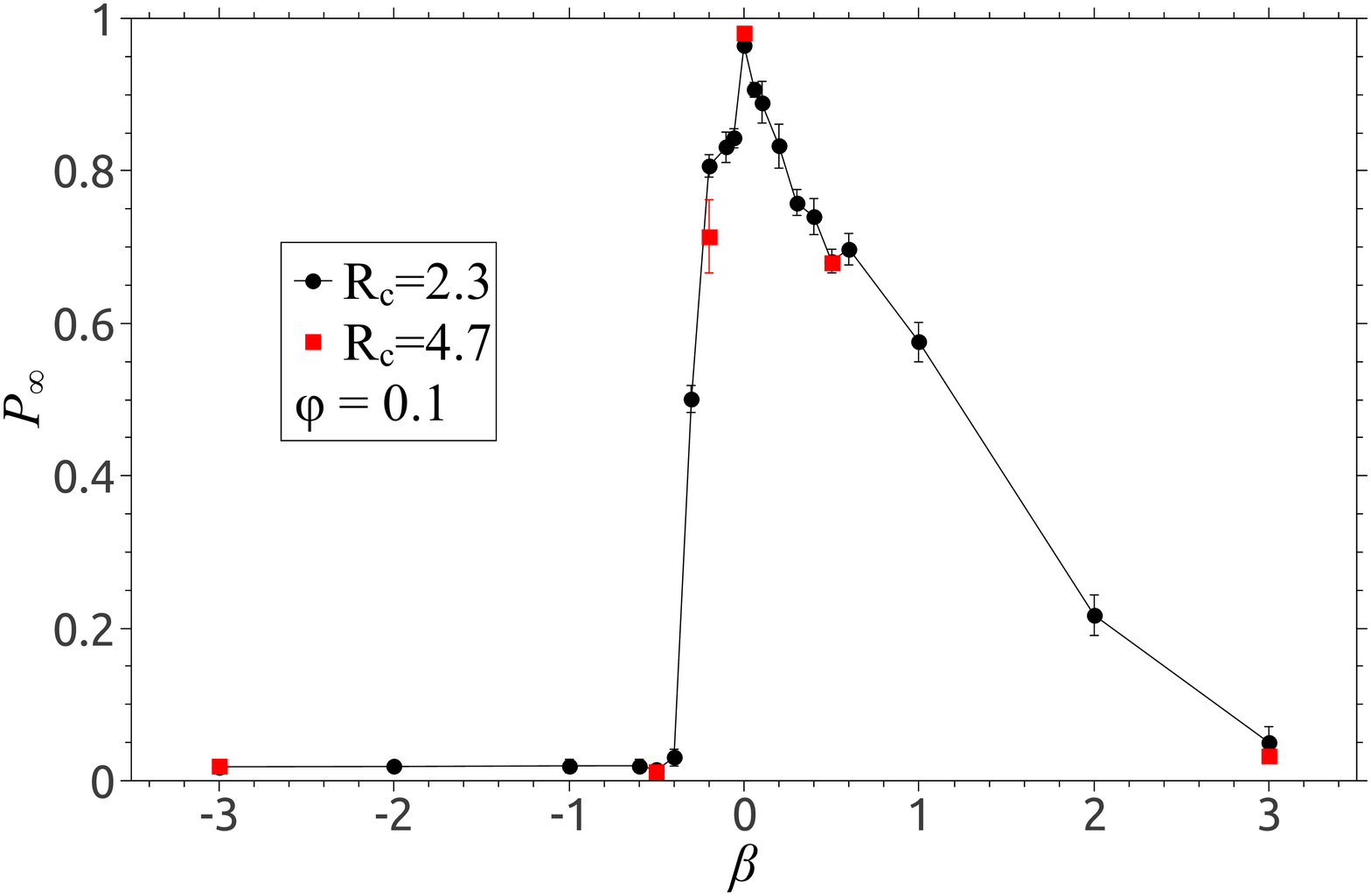}}
%     \includegraphics[scale=0.5]{gr_slide23EPS.pdf}
%    \end{tabular}
\end{center}\vspace{-0.7cm}
  \caption{\small{Long-time Polar Order Parameter, $P_\infty$ for initially aligned suspensions. Results are shown for simulations performed with different squirmer size. The insensitivity of the global order parameter to the squirmer resolution on the simulation lattice indicates that the emergent order and structures described are not controlled by the  details of fluid flow close to the particles. }}\label{P_inf}
\end{figure}

In order to quantify in more detail the role of active stresses in the  global degree of  ordering in squirmer suspensions, we have computed the steady-state value of the polar order parameter, $P_{\infty}$, as a function of the relative apolar stress strength, $\beta$. Fig.~\ref{P_inf} displays  $P_{\infty}$, computed as the mean average of $P(t)$ over the time period after the initial decay from the aligned state~\cite{Lauga}.

%FIGURE%%%%%%
%%%%%%%%%%%%%%%%%
There are two remarkable observations of the results shown in Fig. \ref{P_inf}. First of all, the larger $|\beta|$ the  smaller values of $P_\infty$ observed, which indicate  less squirmer coherence due to  hydrodynamic interactions controlled by the induced active stresses, or  $|\beta|$. Secondly, for a given magnitude of the apolar stress,  $|\beta|$, pullers are more ordered than pushers. Hence,  there is an asymmetry between pullers and pushers. This asymmetry can be explained in terms of the differences in the  near-field interactions between squirmers~\cite{Lauga, Ishikawa_nearfield08}. Squirmer self-propulsion favors head-to-tail collisions~\cite{Ishikawa_Hota_head06} and generates an internal structure that competes with the  tendency of squirmers to rotate due to local flows.  In fact,  head-to-head orientation is stable to rotations for pusher suspensions (as can be clearly appreciated in  the last snapshot of Fig. \ref{Fotos_3Swimmers}, where we can see a lot of pushers interacting head-to-head). In this case, the active stresses  favor head-to-head configurations, which competes with self-propulsion and  decorrelates faster the comoving swimming configurations of squirmers. On the contrary, the stresslet generated by pullers destabilizes  head-to-head configurations favoring the motion  of squirmers along a common director. It is worth noting  that   puller suspensions with $ \beta > 3 $ will evolve to  isotropic configurations, in agreement with the long-time polar order parameter displayed in Fig.~\ref{P_inf}.
   
%FIGURE%%%%%%%FIGURE%%%%%%
\begin{figure}[!ht]
 \begin{center}
%    \begin{tabular}{cc} 
%\resizebox{4.5cm\textit{a} }{!}{\includegraphics{gr_slide23Layer.pdf}}
\scalebox{0.32}{\includegraphics{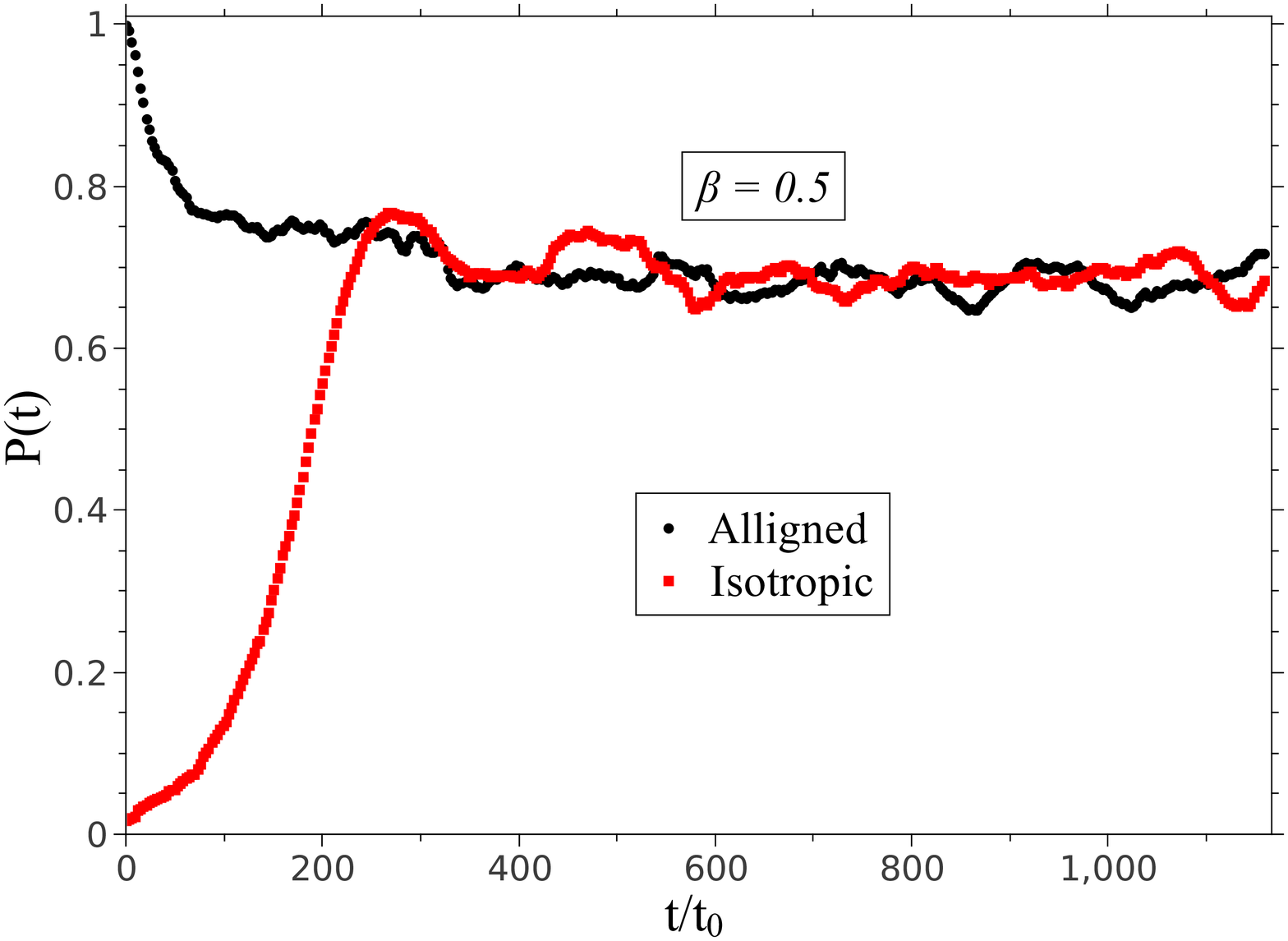}} 
\scalebox{0.57}{\includegraphics{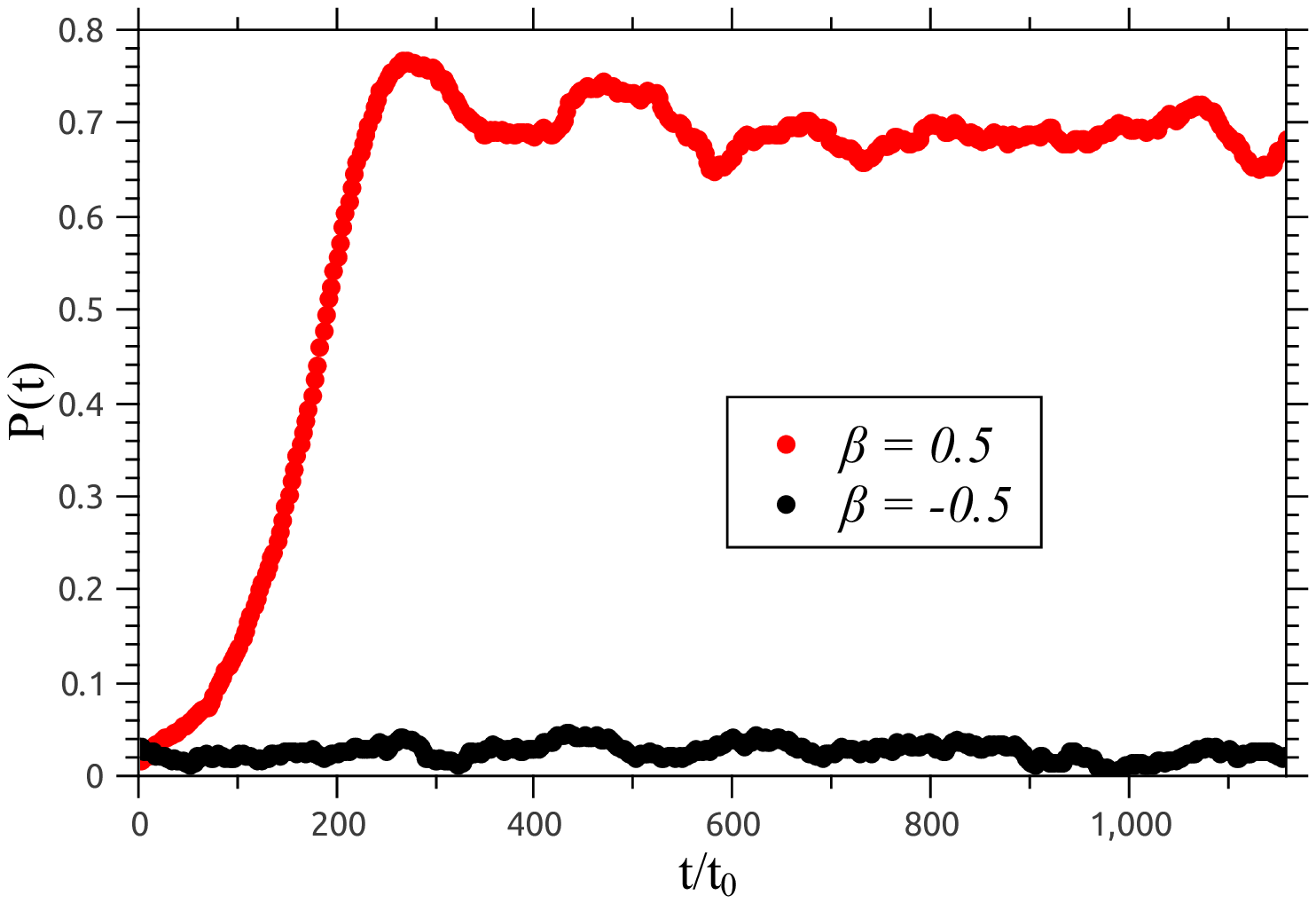}} 
%     \includegraphics[scale=0.5]{gr_slide23EPS.pdf}
%    \end{tabular}
\end{center}\vspace{-0.7cm}
  \caption{Time-evolution of the polar  order parameter, $P(t)$, for squirmer suspensions at $\phi = 1/10$ for different initial configurations. a)  Initially aligned (top) and isotropic (bottom) suspensions of puller squirmers ($\beta = 1/2$). 
  b) Initially isotropic suspensions for completely polar ($\beta = 0$),   puller  ($\beta = 1/2$) and pusher ($\beta = -1/2$) squirmers.}\label{Pt_pullers_both}
\end{figure}

%%%%%%%%%%%%%%%%%%%%%%%%%%%%%%%%
In order to clarify that  global ordering is generic for squirmers composed of spherical particles, and hence that  orientation instabilities do not require   non spherical propelling particles~\cite{Saintillan_08}, we have analyzed the collective evolution of squirmer suspensions with initial   isotropic configurations. It is clear in Fig.~\ref{Pt_pullers_both}.a, that both cases of puller suspensions either initially aligned or isotropic, have a similar long-time polar order; hence we can infer that puller suspensions in  either an  isotropic or aligned state are unstable  and that the steady state is independent of the symmetry of the initial configurations.

%FIGURE%%%%%%%FIGURE%%%%%%
%%%%%%%%%%%%%%%%%%%%%%%%%%%%%%%%%%
On fig.~\ref{Pt_pullers_both}.b one can clearly appreciate  that isotropic  puller suspensions (red circles)  are also unstable, as shown in  Fig.~\ref{Pt_pullers_both}.a. On the contrary,  isotropic pushers suspensions are  stable (black circles) for this regime  of $\beta$.
Similarly to the result for puller suspensions showed in Fig.~\ref{Pt_pullers_both}.a, one can appreciate  in Fig.~\ref{Pt_pullers_both}.b that  pushers are driven to the same long-time polar order parameter, and therefore that the final alignment is independent of the initial configuration.

\section{Flocking.} \label{Patterns_Forms}

\begin{figure}[!ht]
 \begin{center}
%    \begin{tabular}{cc} 
%\resizebox{4.5cm\textit{a} }{!}{\includegraphics{gr_slide23Layer.pdf}}
\scalebox{0.31}{\includegraphics{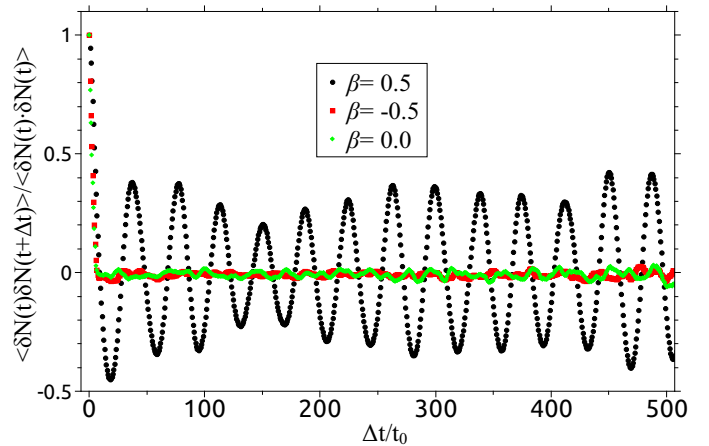}}
%     \includegraphics[scale=0.5]{gr_slide23EPS.pdf}
%    \end{tabular}
\end{center}\vspace{-0.7cm}
  \caption{Temporal correlation functions of density fluctuations}
  \label{fig:flucts}
\end{figure}

\begin{figure}[!ht]
 \begin{center}
%    \begin{tabular}{cc} 
%\resizebox{4.5cm\textit{a} }{!}{\includegraphics{gr_slide23Layer.pdf}}
\scalebox{0.32}{\includegraphics{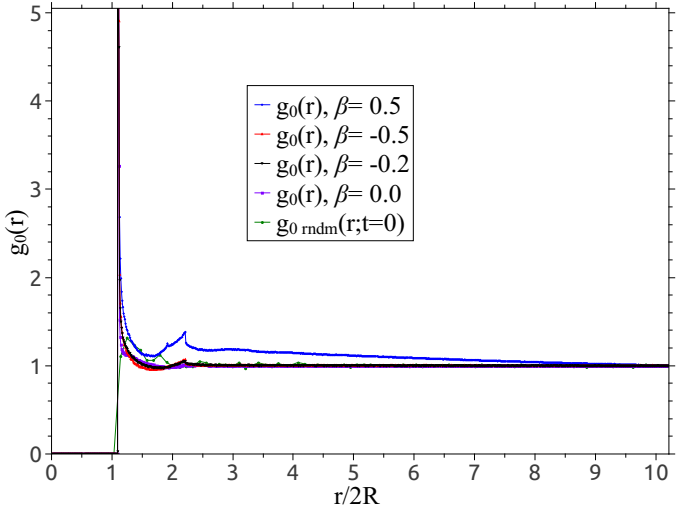}}
%     \includegraphics[scale=0.5]{gr_slide23EPS.pdf}
%    \end{tabular}
\end{center}\vspace{-0.7cm}
  \caption{Radial distribution function, $g_{0}(r)$, for puller ($\beta = 1/2$), pusher ($\beta = -1/2$  and $-1/5$) and totally polar squirmer suspensions  ($\beta = 0$) at $t/t_0 = 870$ time steps. $g_{0\ rndm}(r;t=0)$ is the radial distribution function for the initial configurations where all the squirmers are randomly distributed and completely  aligned.}
  \label{gr_Swimers}
\end{figure}

Fig.~\ref{Fotos_3Swimmers} shows that puller suspensions, $\left( \beta > 0 \right)$,  display a cluster of the size of the box. Due to the absence of attractive forces between squirmers, these observed clusters are statistically relevant but have a dynamic character. As a function of time the observed aggregates evolve and displace; the particles they are form  with change.  We need then a statistical approach  to  analyze the formation of  emergent mesoscale structures  and its correlation with orientational ordering. We have computed the temporal correlation function of the density fluctuations dividing the simulation box in $1000$ sub-boxes of side box $l = L/10$ and counted all the particles $N_i(t)$ at each $i$-th sub-box. This provides the particle temporal mean number, $\langle N_i(t) \rangle_t$,  from which we can determine the instantaneous density fluctuations, $\delta N_i(t) = N_i(t) - \langle N_i(t) \rangle_t$, at each box.  The average density fluctuation, $\delta N(t)$, can then be derived as  the mean of $\delta N_i(t)$ over all the sub-boxes at time $t$, and one can use them to study their temporal correlation. The time correlation of the squirmer density fluctuations, depicted in  Fig.~\ref{fig:flucts}, shows that  pullers have an oscillatory response, associated to the displacement of aggregates with a density  markedly above average, while pushers  are characterized by a more homogeneous spatial distribution. We can gain more  detailed  insight into the aggregation and ordering of squirmer suspensions by studying    the generalized radial distribution functions~\cite{Llopis_epl06}
\begin{equation}\label{gn}
g_n\left(r\right) \equiv \left\langle P_n \left(\cos \theta_{ij}\right)\right\rangle,
\end{equation} 
where $\theta_{ij}$ stands for  the relative angle between the direction of motion of the particles $i$ and   $j$ at a distance between $r$ and $r+dr$ and  $P_n$ is the $n$-th degree Legendre polynomial. For $n = 0$ we recover  the radial distribution function, $g_0\left(r\right)$.  The average in eq. (\ref{gn}) is taken over all  particle pairs and over time, once the system has reached its steady state. Fig.~\ref{gr_Swimers} displays  $g_0\left(r\right)$ for three kinds of squirmers, $\beta = \lbrace 0,1/2,-1/2\rbrace$. For comparison, we also show the radial distribution function of a randomly distributed configuration, which constitutes a good approximation for the equilibrium radial distribution function for hard spheres at $\phi =1/10$.  Fig.~\ref{gr_Swimers} displays also  $g_0\left(r\right)$ for $\beta = -1/5$. This case corresponds to a pusher suspension with the same polar order value, $P_{\infty}$, than the puller suspension at $\beta = 1/2$ and will help to analyze the correlation between global polar order and  the suspension structure.

One can clearly appreciate that activity enhances significantly  the value of the radial distribution at contact, $g_0\left(r = 2R\right)$,  compared with the  corresponding value for an equilibrium suspension. This value is larger for puller suspensions indicating the larger tendency of pullers to remain  closer to each other. The radial distribution function for pullers develops a marked second maximum at $r = 4.25R$ indicating the development of stronger short range structures   for pullers. Neither pushers nor totally polar squirmers have a visible second maximum even when we compare puller and pusher suspensions with equivalent  polar order parameter, $P_{\infty}$. The development of the secondary peak for pullers is consistent with their tendency to form  large aggregates, or flocks,   in agreement with the snapshot depicted in Fig.\ref{Fotos_3Swimmers}.
%FIGURE%%%%%%%FIGURE%%%%%%
%%%%%%%%%%%%%%%%%%%%%%%%%%%

\begin{figure}[!ht]
 \begin{center}
%    \begin{tabular}{cc} 
%\resizebox{4.5cm\textit{a} }{!}{\includegraphics{gr_slide23Layer.pdf}}
\scalebox{0.35}{\includegraphics{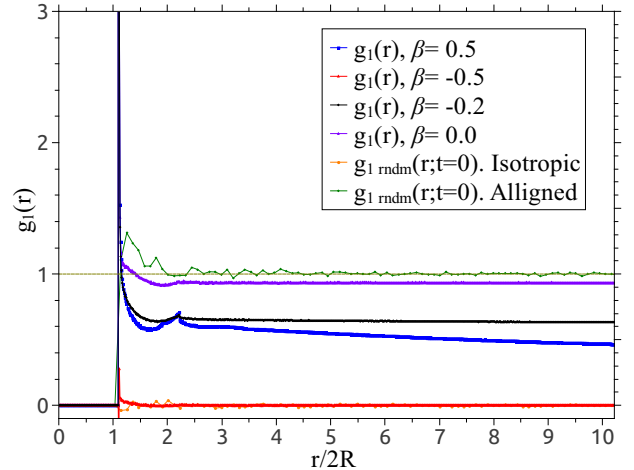}} 
%     \includegraphics[scale=0.5]{gr_slide23EPS.pdf}
%    \end{tabular}
\end{center}\vspace{-0.7cm}
  \caption{$g_{1}(r)$ of pullers ($\beta = 0.5$), pushers ($\beta = -0.5, -0.2$) and totally polar squirmers ($\beta = 0$) at $t/t_0 = 870$, $g_{1\ rndm}(r;t=0)\ Isotropic $ is the correlation function at the beginning of the simulations where all the particles are both at random positions and  orientation. $g_{1\ rndm}(r;t=0)\ Aligned$ is the distribution function at the beginning of the simulations where all the particles are aligned at random positions.}\label{gr1_Swimers}
\end{figure}

Fig.~\ref{gr1_Swimers}  displays the generalized radial distribution function, $g_1(r)$, which provides information on the  degree of local correlated polar order around a given squirmer.  Initially, all squirmers are parallel, and hence $g_1(r, t=0) = 1.0$ (green diamonds in the Figure). The isotropic initial condition  (yellow circles), when $g_1(r,t=0)=0$, is also shown as a reference.
Completely polar squirmer suspensions, $\beta = 0$, keep $g_1(r)$ very close to 1 (violet triangles) showing  that most of the particles swim along  a common  direction even if they are far away from each other; this strong correlation is easily appreciated in the first snapshot of Fig.\ref{Fotos_3Swimmers}. We can observe a similar effect for pusher suspensions at $\beta = -1/5$ where we can see how $g_1(r)$ relaxes to a finite plateau   for $r > 3R$. However,  unlike completely polar  squirmers, now $g_1(r > 3R) \sim 0.6$ (black diamonds)  indicating a loss of coherence in the swimming suspension.  The relative alignment for puller suspensions is clearly different, because $g_1(r)$ decays asymptotically to zero  (blue squares) for separations analogous to those on which  the radial distribution function decays to one. This indicates that the structure we have identified through $g_0(r)$ in Fig.~\ref{gr_Swimers} corresponds to groups of nearby particles that swim along the same direction. This behavior is consistent with the  middle snapshot  of Fig. \ref{Fotos_3Swimmers}  which shows a marked  flocking formed by a significant number of  particles swimming coherently in the same direction. If the apolar  strength is increased, increasing the magnitude of $\beta$,  for  pusher suspensions, the partial coherence that we have seen in the case of $\beta = -1/5$  vanishes.  The curve of $g_1(r)$ for  $\beta = -1/2$ (red triangles) does not display any significant feature, indicating a complete decorrelation in the direction of swimmers at all length scales. The corresponding  configuration in Fig.~\ref{Fotos_3Swimmers} shows  clearly the absence of any significant  correlated orientation between squirmers.

%FIGURE%%%%%%%FIGURE%%%%%%

%%%%%%%%%%%%%%%%%%%%%%%%%
\section{Conclusions}

We have analyzed a model system of swimming spherical particles to show the capabilities of the hydrodynamic coupling as a route to pattern formation, polar ordering and flocking in the absence of any additional interaction among the swimmers (except that swimmers cannot overlap due to excluded volume).  We have shown  how a numerical mesoscopic  model for swimmer suspensions   can  develop  instabilities and long-time polar order and that active stresses play a relevant role to promote flocking due to the coupling of the swimming director with the local fluid vorticity induced by the neighboring squirmers. We have identified the sign of such active stress (which distinguishes pullers from pushers) as the main element which controls  squirmer  flocking and swimming coherence.

We have shown that spherical squirmers, starting from aligned or isotropic state, develop a unique long-time polar order due to  hydrodynamic interactions. We have found that aligned pushers suspension are unstable while  isotropic suspensions  are stable for $\beta < -2/5$: isotropic puller suspensions are also stable for  $\beta > 3.0 $.  

We have seen that flocking configurations for pullers leads to large elongated structures, reminiscent of the bands observed in the  Vicsek model~\cite{Vicsek}. However, in this later case hydrodynamics is absent and flocking develops at high concentrations, when the aligning interaction is strong  enough to overcome decoherence induced by noise. In the systems we have explored the coherence is hydrodynamic and develops at small volume fractions.  The observed elongated, spanning aggregates with internal coherent orientation, in the range   $0 < \beta < 1$, are robust and independent of  the initial  configuration.

%%%%%%%%%%%%%%%%%%%%%%%%%%%%%%%%%%%%%%%%%%%%%%%%%%%
\section*{Acknowledgements}
The authors acknowledge R. Matas-Navarro and A. Scagliarini for useful discussions. We acknowledge MINECO (Spain) and DURSI  for financial support under Projects No. FIS2011-22603 and No. 2009SGR-634, respectively. F.A. acknowledges support from Conacyt (Mexico). The computational work herein was carried out in the MareNostrum Supercomputer at Barcelona Supercomputing Center.

%%%%%%%%%%%%%%%%%%%%%%%%%%%%%%%%%%%%%%%%%%%%%%%%%%%%%%%
% A useful Journal macro
\def\jour#1#2#3#4{{#1} {\bf #2}, #3 (#4).}
\def\tbp#1{{\em #1}, to be published.}
\def\tit#1#2#3#4#5{{#1} {\bf #2}, #3 (#4).}
\def\ap{Adv. Phys.}
\def\epl{Europhys. Lett.}
\def\epjB{Eur. Phys. J. B}
\def\epjE{Eur. Phys. J. E}
\def\cpam{Comm. Pure Appl. Math.}
\def\prl{Phys. Rev. Lett.}
\def\pr{Phys. Rev.}
\def\pra{Phys. Rev. A}
\def\prb{Phys. Rev. B}
\def\pre{Phys. Rev. E}
\def\pa{Physica A}
\def\ps{Physica Scripta}
\def\pf{Phys. Fluids}
\def\jmpc{J. Mod. Phys. C}
\def\jpc{J. Phys. C}
\def\jcp{J. Chem. Phys.}
\def\jpcs{J. Phys. Chem. Solids}
\def\jpco{J. Phys. Cond. Mat}
\def\jf{J. Fluids}
\def\jfm{J. Fluid Mech.}
\def\arf{Ann. Rev. Fluid Mech.}
\def\roy{Proc. Roy. Soc.}
\def\rmp{Rev. Mod. Phys.}
\def\jsp{J. Stat. Phys.}
\def\pla{Phys. Lett. A}
%%%%%%%%%%%%%%%%%%%%%%%%%%%%%%%%%%%%%%%%%%%%%%%%%%%%%

\end{document}